\documentclass[fleqn,twoside]{article}
\usepackage{espcrc2}

\usepackage{graphicx}

\newcommand{\AmS}{{\protect\the\textfont2
  A\kern-.1667em\lower.5ex\hbox{M}\kern-.125emS}}

\title{$B$ Physics and Supersymmetry}

\author{L. Silvestrini\address[MCSD]{INFN, Sez. di Roma, Dip. di
    Fisica, Univ. di Roma ``La Sapienza'',\\ 
        P.le A. Moro 2, I-00185 Roma, Italy}}
       
\begin{document}

\begin{abstract}
In this talk, I briefly review a selection of SUSY effects in $B$
physics. First, I consider models with Minimal Flavour Violation. Then
I discuss SUSY models with new sources of flavour violation in
squark mass matrices, analyzing present constraints and possible
developments with forthcoming data on $b\to s$ and $b \to d$ transitions.
\vspace{1pc}
\end{abstract}

\maketitle

\section{INTRODUCTION}
\label{sec:intro}

Flavour changing neutral currents (FCNC) and CP-violating processes
are very sensitive probes of new physics. The GIM suppression of FCNC
amplitudes is generally absent beyond the Standard Model (SM), giving
possibly large enhancements of FCNC processes over the SM predictions.
Furthermore, CP violation in weak decays in the SM is entirely
governed by a single phase in the CKM matrix, resulting in a strong
correlation between all CP-violating processes. In general, this
correlation is lost beyond the SM. Until now, all experimental data
are fully compatible with the SM, as can be clearly seen from the
standard analysis of the Unitarity Triangle (UT) \cite{UT}. One is then
left with two basic questions: \textit{i)} How can the SM be extended
without spoiling the impressive consistency of the SM UT fit?
\textit{ii)} Can we still hope to see some hint of new physics in
low-energy FCNC and CP-violating phenomena? While these two questions
can be considered in full generality, I will focus in the rest of this
talk on Supersymmetry (SUSY), which is at present the most successful
extension of the SM with respect to the consistency with precision
electroweak data. SUSY can modify the SM predictions on FCNC and CP-violating
processes in three different ways:
1) With additional loop contributions still proportional to
  elements of the CKM matrix. A typical example of this kind of
  contributions is given by stop and chargino loops. This effect is
  present in any SUSY extension of the SM.
2) With additional loop contributions governed by new sources of
  flavour and CP violation \cite{Hall}. When these new sources of FCNC are
  present, typically the largest contributions arise from
  gluino exchange. This effect is absent in SUSY models with
  Minimal Flavour Violation (MFV) (see Sec.~\ref{sec:mfv}).
3) With additional tree-level contributions. These can only arise
  in models with R-parity violation, and typically affect FCNC
  processes in a dramatic way.

Clearly, models in which only the first kind of contributions are
present tend to agree much better with experimental data than
more general models. On the other hand, the presence of contributions
of the second or third kind usually generates larger deviations from
SM predictions on yet unobserved FCNC and CP violating processes.
In this respect, $B$ physics is the next frontier of testing SUSY in
weak decays: $B$ factories and the Tevatron will provide us with data
on a large variety of FCNC and CP-violating $B$ decays, and $B$
physics is playing an ever-increasing role in the UT fit. In the
following, I will discuss in some detail SUSY effects in $B$ physics
in models with R-parity conservation.\footnote{For a recent example of the
  large effects in $B$ physics generated by R-violating couplings, see
  \cite{Bar-Shalom:2002sv}.}

\section{$B$ PHYSICS IN MODELS WITH MFV}
\label{sec:mfv}

Let me start by considering models with MFV, defined by the following
requirements. First, I assume that all sfermion masses are flavour
diagonal and real at the electroweak scale, and all gaugino and Higgs
mass parameters are real. This ensures that the CKM matrix is the only
source of flavour and CP violation, so that only the first kind of
contributions discussed above can arise. Second, I assume that $\tan
\beta$ is not too large ($\tan \beta \le 10$), so that no new operator
is generated in the effective Hamiltonian for $B$ decays. This implies
that the SM analysis of perturbative and non-perturbative strong
interaction effects can also be applied in this model. Finally, I
assume that all squarks are degenerate except for the stop. This
ensures that SUSY only modifies the contribution of the top quark in
the SM. In particular, this means that SUSY contributions cancel from
all quantities that in the SM do not depend on the top quark mass. A
typical example is given by the ratio of the mass difference of $B_d$
and $B_s$ mesons. Indeed, a Universal Unitarity Triangle (UUT) can be
constructed in these models, independent of SUSY contributions
\cite{UUT}, once a sufficient number of top-mass independent
quantities becomes available. At present, this is not the case and one
needs to perform the full UT analysis to determine the CKM parameters
for any given value of SUSY masses. The relevant SUSY parameters for
the analysis are stop masses and mixing angle, the mass of the
lightest chargino, the $\mu$ parameter, $\tan \beta$ and the charged
Higgs mass. Potentially large contributions to $b \to s \gamma$ arise;
once the constraints from $b \to s \gamma$ and the lower bounds on
Higgs and SUSY masses from direct searches are taken into account, the
UT fit in these models is indistinguishable from the SM one
\cite{FCNCMFV,Buras}. Deviations at the level of $20-30 \%$ are
possible in rare $B$ decays \cite{Buras}. This is a property shared also
by non-SUSY MFV models \cite{Parodi}. In the near future, $b \to s
\gamma$ and $b \to s \ell^+ \ell^-$ will be the most sensitive probes
of SUSY models with MFV \cite{Strumia}. 

If one allows $\tan \beta$ to be very large, huge effects are possible
in $B_s \to \mu^+ \mu^-$ \cite{bsmumu}, increasing the BR from $BR(B_s
\to \mu^+ \mu^-)_{SM}\sim 4 \cdot 10^{-9}$ to $10^{-6}$, behind the
corner of $BR(B_s \to \mu^+ \mu^-)_{EXP}<2 \cdot 10^{-6}$
\cite{Tevatron}. This large enhancement of $B_s \to \mu^+ \mu^-$ is in
general correlated to a decrease of the $B_s - \bar B_s$ mass
difference $\Delta M_s$, cutting out a large part of parameter space
\cite{polacchi}. Once a measurement of $\Delta M_s$ becomes available,
this correlation will be very useful in determining the prediction for
$B_s \to \mu^+ \mu^-$ in this model.  It is also interesting to notice
that, in minimal supergravity models, a correlation can be established
between contributions to $B_s \to \mu^+ \mu^-$ and $(g-2)_\mu$. In the
region of parameter space favoured by the anomalous magnetic moment of
the muon, one can obtain enhancements of one or two orders of
magnitude of $BR(B_s \to \mu^+ \mu^-)$ over the SM prediction \cite{Dedes}.

\section{$B$ PHYSICS IN GENERAL SUSY}
\label{sec:general}

\begin{figure*}
\begin{center}
\begin{tabular}{c c}
\includegraphics[width=7.5cm]{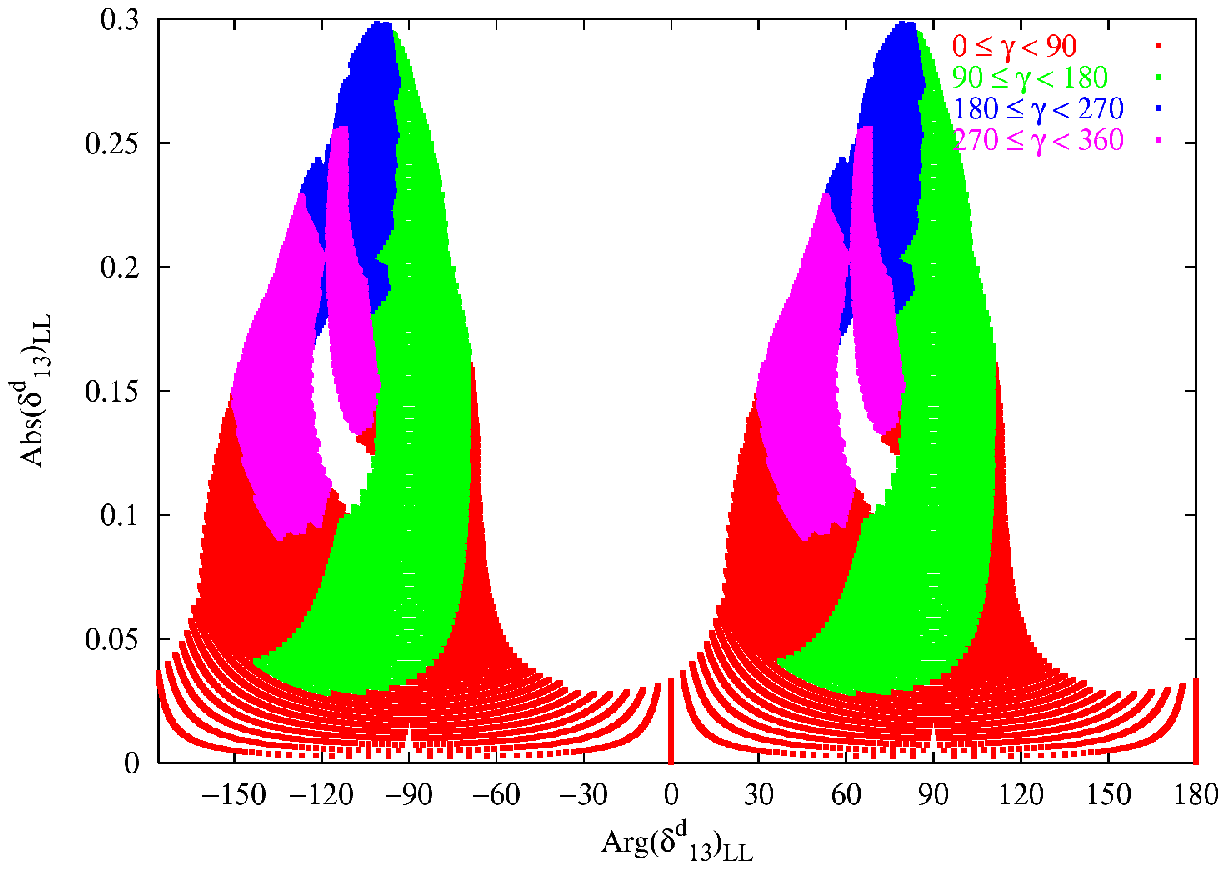} &
\includegraphics[width=7.5cm]{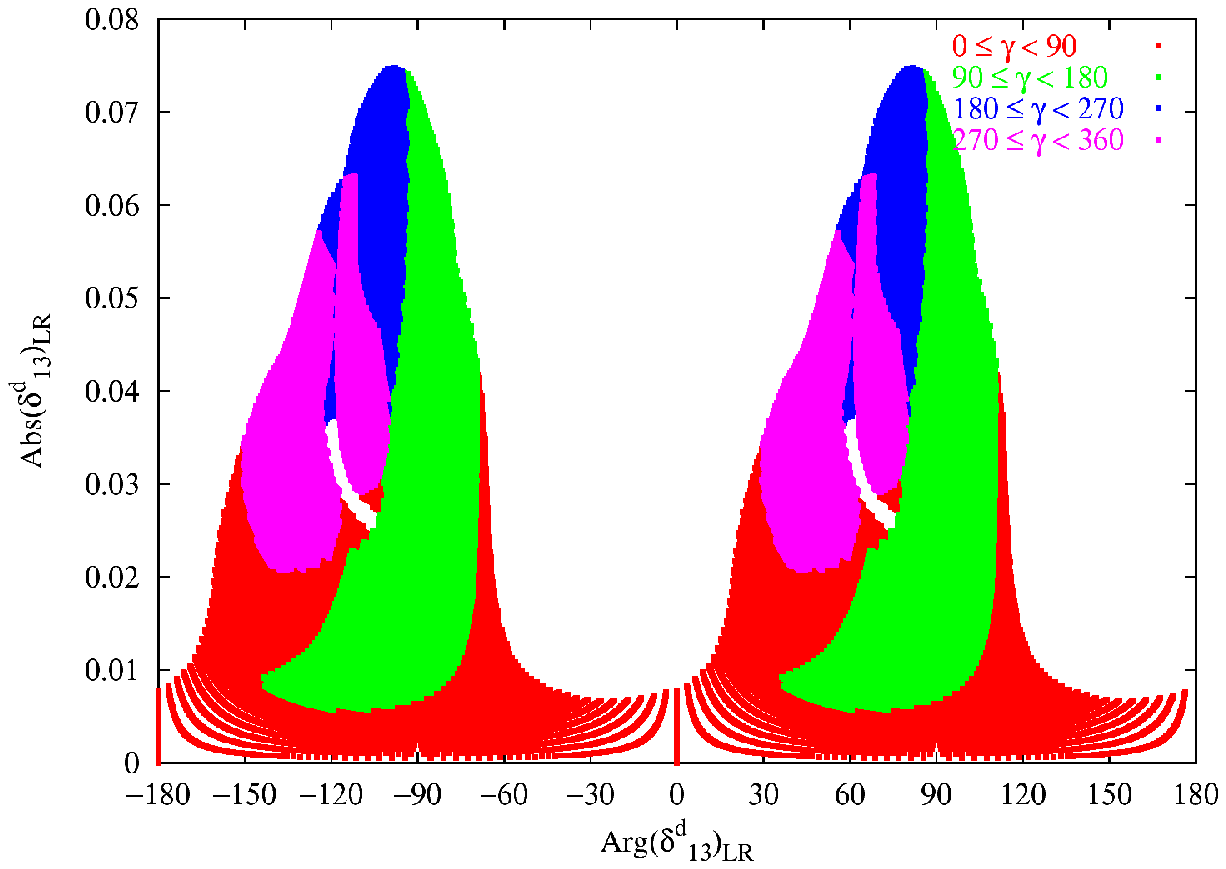} \\
\includegraphics[width=7.5cm]{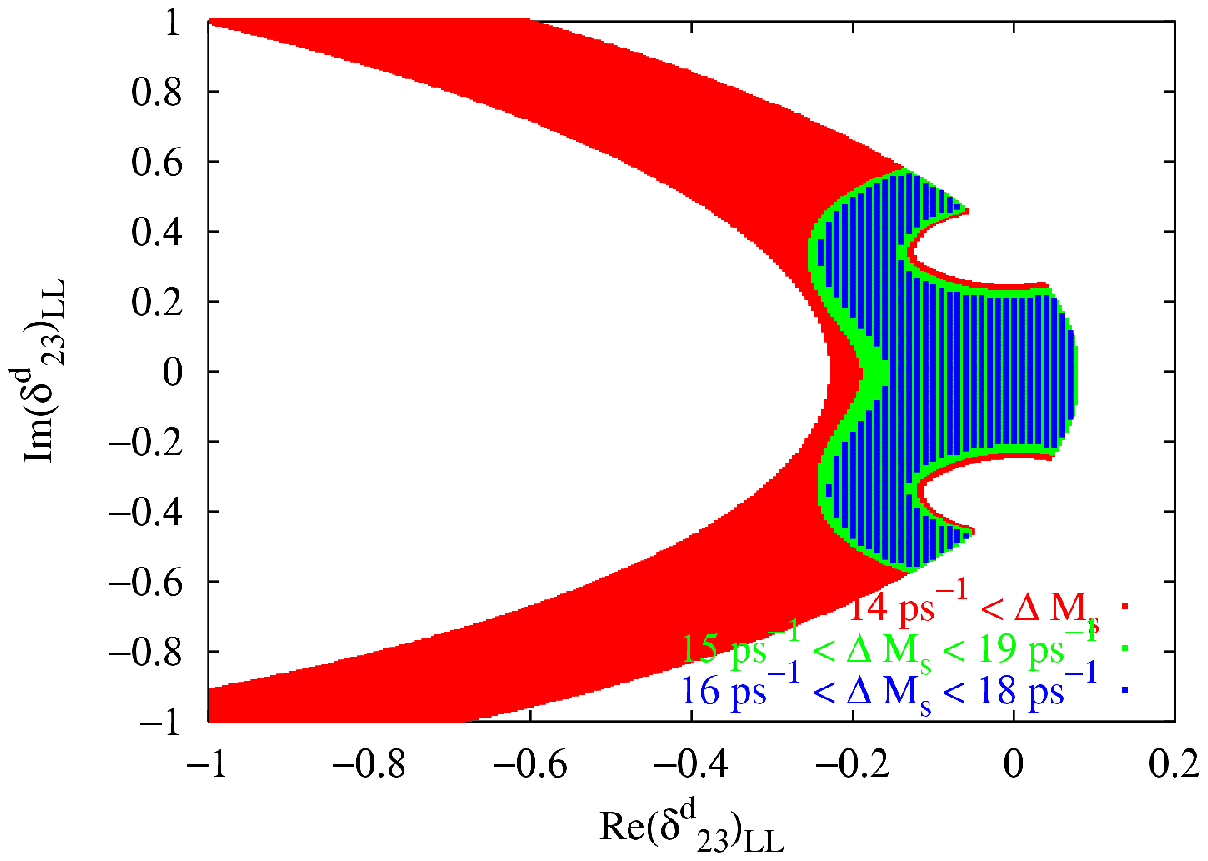} &
\includegraphics[width=7.5cm]{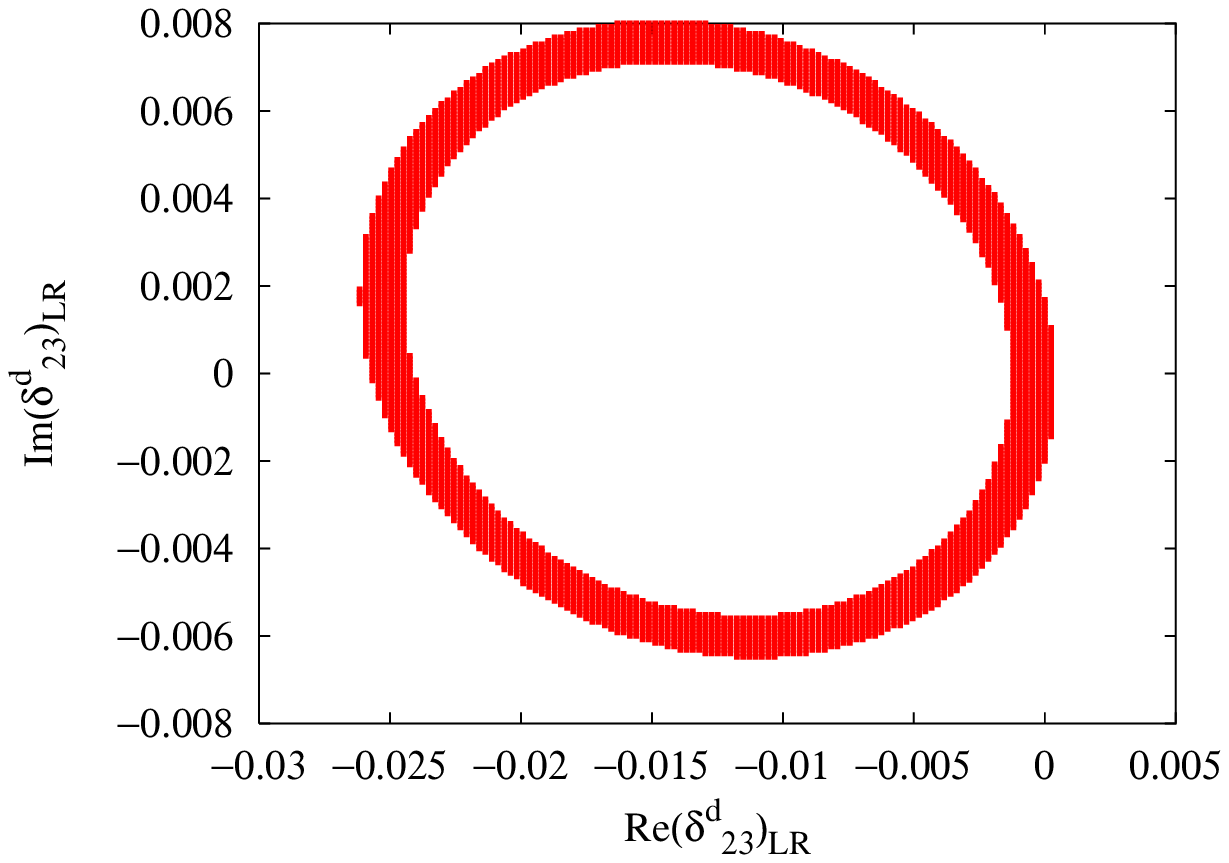} \\
\includegraphics[width=7.5cm]{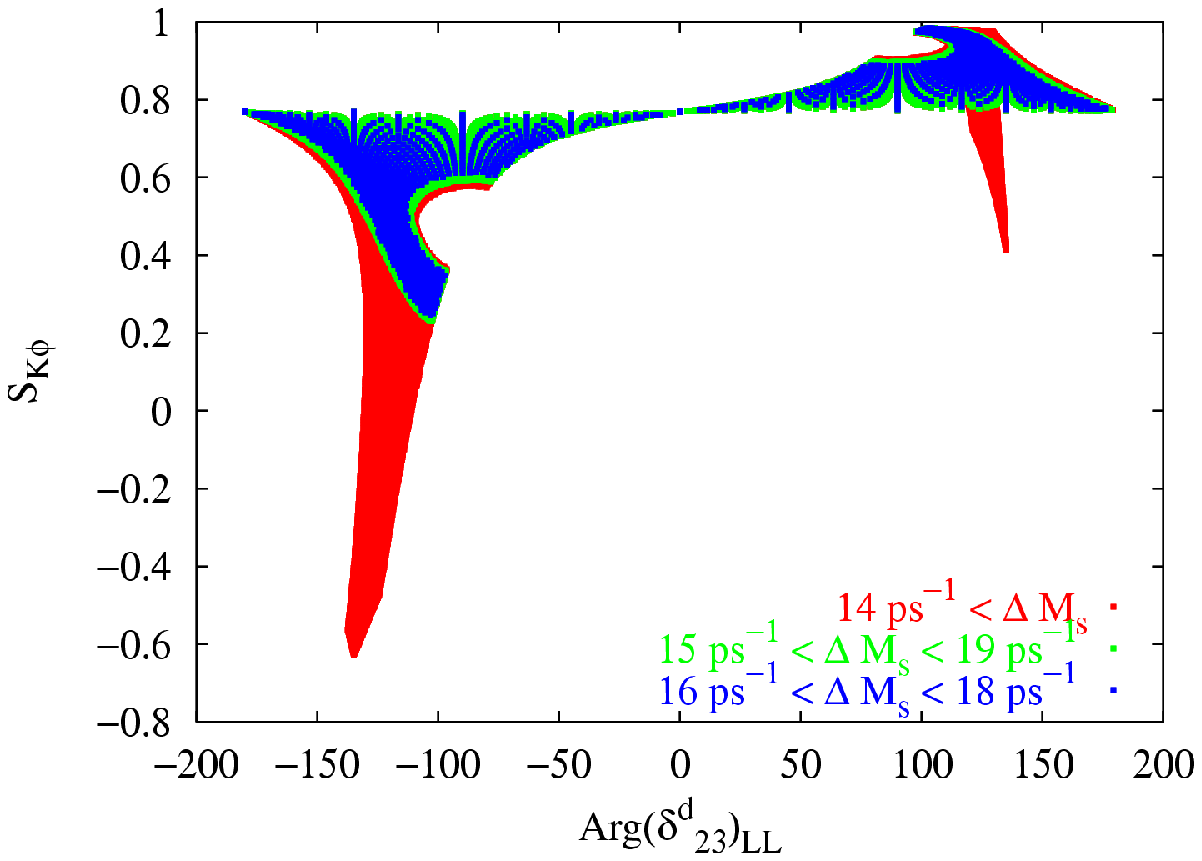} &
\includegraphics[width=7.5cm]{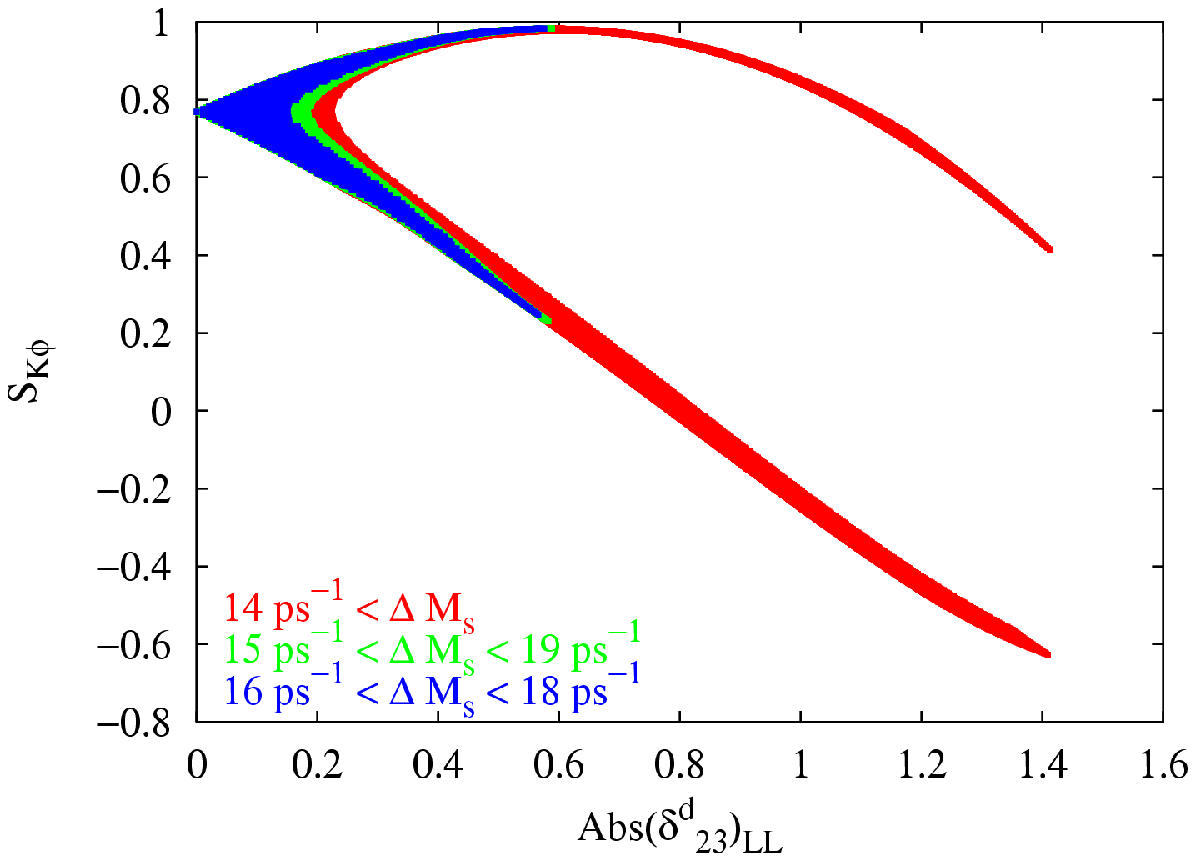} \\
\end{tabular}
\end{center}
\caption{Top: Allowed values of Abs$(\delta^{d}_{13})_{AB}$ as a
  function of Arg$(\delta^{d}_{13})_{AB}$ for $AB=LL,LR$ and
  $m_{\tilde q}=500$ GeV (from~\cite{Becirevic:2001jj}). Different
  colours denote values of $\gamma$ belonging to different quadrants.
  Middle: Allowed regions in the Re$(\delta^{d}_{23})_{AB}$ --
  Im$(\delta^{d}_{23})_{AB}$ plane for $AB=LL,LR$ and $m_{\tilde
    q}=250$ GeV (from~\cite{noifuture}). Results are preliminary.
  Bottom: $S_{\phi K_s}$ in the presence of $(\delta^{d}_{23})_{LL}$
  for $m_{\tilde q}=250$ GeV (from~\cite{noifuture}).  Results are
  preliminary. }
\label{fig:abs}
\end{figure*}

We now turn from MFV to SUSY models with arbitrary sfermion mass
matrices. In general, FCNC and CP-violating processes impose stringent
constraints on off-diagonal sfermion mass terms \cite{GMS}.  To study
these models in full generality, it is convenient to parameterize FCNC
amplitudes in terms of $(\delta^d_{ij})_{AB}$, the ratio of the
off-diagonal squark mass term $(\Delta^d_{ij})_{AB}$ connecting
squarks of flavour $i$ and $j$ and helicities $A$ and $B$ over the
average squark mass $m^2_{\tilde q}$. Let us first consider
$(\delta^d_{13})_{AB}$, the mass insertion that induces $b
\leftrightarrow d$ transitions. Constraints on this parameter from the
$B_d - \bar B_d$ mass difference $\Delta M_d$ and from the
time-dependent CP asymmetry in $B_d \to J/\Psi K$ decays $a_{\Psi K}$
have been recently studied in~\cite{Becirevic:2001jj}. This analysis
includes NLO QCD corrections \cite{QCDDB2} and lattice matrix elements
\cite{lattdb2}. As an example of the constraints one obtains from this
kind of analysis, I report in Fig. \ref{fig:abs} the allowed regions
in the Abs$(\delta^{d}_{13})_{AB}$ -- Arg$(\delta^{d}_{13})_{AB}$
plane $(AB=LL,LR)$, for $m_{\tilde q}=500$ GeV.  Other results and
details of the analysis can be found in~\cite{Becirevic:2001jj}. A
similar analysis including also chargino contributions has been very
recently carried out in~\cite{Gabrielli:2002fr}.  Effects of
$(\delta^d_{13})_{LR}$ can also be tested using $B \to \rho \gamma$
decays \cite{Ali:2002kw}.

A similar analysis can be carried out for $(\delta^{d}_{23})_{AB}$,
the mass insertions that generate $b \leftrightarrow s$ transitions
\cite{noifuture}. In this case, however, at present we only have a
lower bound on $\Delta M_s$ and the precise measurement of $BR(B \to
X_s \gamma)$. The latter quantity is very effective in constraining
$(\delta^{d}_{23})_{LR}$, while its impact on $(\delta^{d}_{23})_{LL}$
is quite limited. In Fig. \ref{fig:abs} I report preliminary results
on the allowed regions in the Re$(\delta^{d}_{23})_{AB}$ --
Im$(\delta^{d}_{23})_{AB}$ plane, for $AB=LL,LR$, for the present
lower bound $\Delta M_s > 14$ ps$^{-1}$. To illustrate the possible
impact of a future measurement of $\Delta M_s$, in Fig.
\ref{fig:abs} I also report the allowed regions for $15$ ps$^{-1} <
\Delta M_s < 19$ ps$^{-1}$ and $16$ ps$^{-1} < \Delta M_s < 18$
ps$^{-1}$. While the constraints on $(\delta^{d}_{23})_{LR}$ are
completely dominated by $BR(B \to X_s \gamma)$, one can see clearly
that a measurement of $\Delta M_s$ will drastically reduce the allowed
values of $(\delta^{d}_{23})_{LL}$. However, given the errors on
hadronic matrix elements, the allowed range for
$(\delta^{d}_{23})_{LL}$ cannot shrink too much when the
experimental error on $\Delta M_s$ is reduced.

Clearly, $(\delta^{d}_{23})_{AB}$ enter many other interesting $b \to
s$ decays. Let us first consider $B \to \phi K_s$. The first
measurements of $a_{\phi K_s}$ by the BaBar and Belle collaborations
display a $2.7 \sigma$ deviation from the observed value of $a_{\Psi
  K}$ \cite{exps2b}, while in the SM both quantities should measure
$\sin 2 \beta$ with negligible hadronic uncertainties
\cite{Grossman:1997gr} (here $\beta$ is one of the angles of the UT).
In Fig. \ref{fig:abs} I report $S_{\phi K_s}$ (the coefficient of the
$\sin \Delta M_d t$ term in $a_{\phi K_s}$) as a function of
$(\delta^{d}_{23})_{LL}$ for different values of $\Delta M_s$
\cite{noifuture}. At present, SUSY effects can account for the
observed central value, but this may change with future data on
$\Delta M_s$. It is interesting to notice that direct CP violation can
also occur in this channel \cite{instant}.  Another interesting
process in which effects of $(\delta^{d}_{23})_{AB}$ can be seen is $b
\to s \ell^+ \ell^-$. Here large deviations from the SM in the
asymmetries can be generated even for values of the BR close to the SM
expectations \cite{bsll}.

\section{CONCLUSIONS}
\label{sec:concl}

For reasons of space, in this talk I was able to discuss only a few of
the many interesting implications of SUSY in $B$ physics. However,
already from these selected topics it is clear that the richer the
flavour structure of superpartners is, the more probable it is to
discover indirect signs of SUSY in $B$ physics. In any case,
forthcoming data in this field will certainly help us learn more on
SUSY extensions of the SM.


\subsection*{ACKNOWLEDGMENTS}
I am much indebted to M. Ciuchini for his
kind and patient help in preparing the figures.

\end{document}